# The gravitational heat conduction and the hierarchical structure in solar interior

Zheng Yahui[1,2] and Du Jiulin[1]

1. *Department of Physics, School of Science, Tianjin University, Tianjin 300072, China*
2. *Department of Physics, School of Science, Qiqihar University, Qiqihar City 161006, China.*

**Abstract** With the assumption of local Tsallis equilibrium, the newly defined gravitational temperature is calculated in the solar interior, whose distribution curve can be divided into three parts, the solar core region, radiation region and convection region, in excellent agreement with the solar hierarchical structure. By generalizing the Fourier's law, one new mechanism of heat conduction, based on the gradient of the gravitational temperature, is introduced into the astrophysical system. This mechanism is related to the self-gravity of such self-gravitating system whose characteristic scale is large enough. It perhaps plays an important role in the astrophysical system which, in the solar interior, leads to the heat accumulation at the bottom of the convection layer and then motivates the convection motion.

**Keywords:** Gravitational temperature; heat conduction; convection motion.

**PACS:** 44.10.+i; 51.20.+d; 96.60.Jw.

As everyone knows, almost all the physical systems treated in statistical mechanics with the Boltzmann–Gibbs (BG) entropy are extensive and this property holds for the systems with short-range interactions. However, when dealing with the systems with long-range interactions such as Newtonian gravitational forces and Coulomb electric forces, the BG statistics may need to be generalized for their statistical description.

In recent years, nonextensive statistics mechanics (NSM) based on Tsallis entropy [1] has been proposed as a generalization of BG statistics. This new statistics theory has been used to deal with many interesting problems in the systems with long-range interactions, among which the self-gravitating system is one important research field [2–6]. The theoretic and observational successes for the NSM in this field show that, possibly this statistics theory is one fundamentally valid method to describe the astrophysical system. Combining the Newtonian mechanics or general relativity, it can be used to interpret various phenomena in astrophysics.

On the other hand, it is difficult for the scientists to carry out the thermodynamic research on the astrophysical system. The ordinary results they could obtain in their researches are the gravothermal catastrophe [7,8] and the negative thermal capacity [7]. These results prevent people from getting more insight into the essence of the gravity in the view of thermodynamics. Although with the NSM these results can be avoided partly [9], it is very difficult to formally introduce the heat conduction mechanism, in the astrophysical sense, into the self-gravitating system. The main reason for this is that, in the development process of the NSM, the controversy on the temperature definition holds back the emergence of the thermodynamics in the framework of the NSM.

Therefore, the key to fill in the gaps in the astrophysical thermodynamics is to present one feasible idea about the temperature definition. Perhaps it can be done through the assumption of temperature duality [10], according to which the gravitational temperature can de defined in the astrophysical systems [11]. With the gravitational temperature, the heat conduction driven by the gravity can be naturally discovered. Before this, let us firstly introduce the generalized Maxwellian velocity distribution function [12,13] into the self-gravitating system,

$$f_q(\vec{r},\vec{v}) = nB_q (\frac{m}{2\pi kT})^{\frac{3}{2}} [1-(1-q)\frac{mv^2}{2kT}]^{\frac{1}{1-q}} \quad (1)$$

where $n$ is the number density, $m$ is the particle's average mass, $T$ is the thermodynamic temperature (the inverse of Lagrange multiplier, or Lagrange temperature), and $B_q$ is the normalized constant related to the nonextensive parameter $q$. This distribution function is the solution of the generalized Boltzmann equation [13] when the system is at the Tsallis equilibrium state.

People always believe that one self-gravitating system at the stable state (there is no nuclear reaction in its center and there is also not convection taking place inside the system), is ordinarily at the hydrostatic equilibrium, so it satisfies the hydrostatic equation

$$\nabla P = -mn\nabla \varphi \quad (2)$$

where the $\varphi$ is the gravitational potential of the system. Now that the temperature in the whole system is inhomogeneous, in the view of thermodynamics, the self-gravitating system should be also at the nonequilibrium stationary state. According to Eqs.(1) and (2), one can realize that the Tsallis equilibrium state is identical to the nonequilibrium stationary state in such systems (On the hydrostatic equilibrium and Tsallis equilibrium for self-gravitating systems, the reader can see Ref.[14]). Just in this sense, the gravitational temperature is defined.

One of the important results from Eq.(1) and the $q$-H theorem is the equation [2],

$$k\nabla T + (1-q)m\nabla\varphi = 0, \quad (3)$$

where $Q=1-q$. Eq.(3) describes the relationship between the Lagrange temperature gradient and gravitational field strength, and it also states the fact that the Lagrange temperature in the astrophysical systems is ordinarily inhomogeneous. Now that the system is at Tsallis equilibrium when Eqs.(1) and (2) are both satisfied, there should not be any heat conduction in such system under this condition. Otherwise, the principle of maximum entropy will be violated. This is difficult to understand, because according to the Fourier's law the temperature gradient always leads to the heat conduction.

It can be explained through Fig 1. Considering one self-gravitational gaseous sphere, A and B are two spherical shells which contain the same amount of molecules. Setting the distance $L$ between shell A and shell B as the free path length at the corresponding part, and when the molecules with high average kinetic energy move from A to B due to thermal collisions, their kinetic energies will decrease because of the work done by the self-gravity of the system. When they arrive at the shell B they can not contribute to the average kinetic energy in B under certain condition. Similarly, when the molecules move from B to A, they also can not change the average kinetic energy in A. That is to say, when these two shells exchange the same number of molecules, there would be no net energy flux between them, although the Lagrange temperature in A is higher than the one in B. Of course, here the Lagrange temperature is related to the molecular average kinetic energy.

On the other hand, there is one mechanism to keep the equal amount exchange between A and B. The molecules in shell B tend to move to A under the influence of self-gravity, and due to the higher density and pressure in A than B, the molecules in A also tend to B. These two trends can balance each other under certain condition, that is to say, these two shells can exchange the same number of molecules in same time. In the above analysis, the influence of thermal radiation is neglected. When considering the effect of thermal radiation, the situation would be more complicated. Here we would not give more analysis.

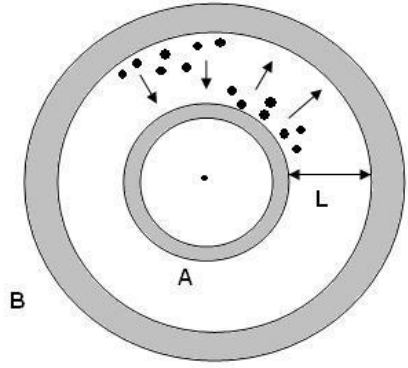

Fig 1. The schematic diagram of no heat conduction in a self-gravitating gaseous sphere at Tsallis equilibrium state.

One can know that, due to the existence of self-gravity, one self-gravitational system can keep the Lagrange temperature gradient for long time. The above analysis indicates that, in Eq.(3), the heat flux determined by the Lagrange temperature gradient and the energy flow determined by the gravitational potential gradient can cancel each other under the condition of Tsallis equilibrium. That is to say, Eq.(3) determine the state of the whole self-gravitating system. So, the most important state parameter gravitational temperature can be defined according to them. That is [11],

$$T_g = T + \frac{Qm\varphi}{k} + T_0. \qquad (4)$$

The temperature constant $T_0$ is determined by the boundary condition.

Here, we regard condition Eq.(2) as the condition of Tsallis equilibrium. When Eq.(2) is satisfied, the system is at the Tsallis equilibrium state, and there must be $\nabla T_g=0$, so Eq.(3) is satisfied. On the contrary, when Eq.(2) is not satisfied, the whole system is not at Tsallis equilibrium state, so there must be $\nabla T_g \neq 0$ in the whole system, under which condition, Eq.(3) can not hold. However, we can generalize the assumption of local equilibrium: when the whole system is not at the Tsallis equilibrium, in local sense, the corresponding part of the system could be at the Tsallis equilibrium, so Eqs.(2) and (3) both hold in this part.

In other words, no matter whether the parameter $Q$ is homogeneous or not, Eq.(3) always holds. This indicates that Eq.(3) is related to the hydrostatic equation Eq.(2): as long as the system is at hydrostatic equilibrium Eq.(3) must hold, although the parameter $Q$ in the equation is possibly different in different part of the system. So Tsallis equilibrium state is not always identical to the nonequilibrium stationary state.

When the self-gravitating system is not at the Tsallis equilibrium, with the hydrostatic equilibrium and according to the analysis in Fig 1, we know that there would be heat conduction in the system. The heat flux can be defined through the generalization of Fourier law:

$$\bar{J}_g = -k_g \nabla T_g. \tag{5}$$

Obviously, the heat flux vector in the above law is dependent on the gradient of gravitational temperature; therefore it can be called gravitational heat flux. Correspondingly, the parameter $k_g$ in Eq.(5) is the gravitational heat conduction coefficient whose value should be positive, and the heat conduction process is gravitational heat conduction.

| $r/R_\odot$ | $T$ | $dT/dr$ | $\dfrac{M_r/M_\odot}{(r/R_\odot)^2}$ | $-\dfrac{M_r/M_\odot}{r/R_\odot}$ | $T_g$ |
|---|---|---|---|---|---|
| 0.02020 | 1.53E+07 | −9615384.6 | 2.1201 | −5.0006 | −1.97E+06 |
| 0.06291 | 1.43E+07 | −28721876.5 | 5.7480 | −4.8284 | −4.43E+06 |
| 0.10113 | 1.29E+07 | −37960954.5 | 7.6198 | −4.5700 | −4.46E+06 |
| 0.18051 | 9.92E+06 | −35599194.4 | 8.5712 | −3.9091 | −9.45E+05 |
| 0.22107 | 8.63E+06 | −30620347.4 | 8.2036 | −3.5680 | 6.79E+05 |
| 0.30165 | 6.65E+06 | −21253406.0 | 6.7928 | −2.9607 | 2.75E+06 |
| 0.38169 | 5.27E+06 | −15073170.7 | 5.2818 | −2.4786 | 3.57E+06 |
| 0.42236 | 4.73E+06 | −12888688.6 | 4.6149 | −2.2775 | 3.74E+06 |
| 0.46071 | 4.29E+06 | −11194029.9 | 4.0649 | −2.1113 | 3.84E+06 |
| 0.50184 | 3.87E+06 | −9851988.9 | 3.5566 | −1.9548 | 3.82E+06 |
| 0.58342 | 3.16E+06 | −8101742.8 | 2.7622 | −1.6987 | 3.55E+06 |
| 0.62194 | 2.86E+06 | −7661085.7 | 2.4676 | −1.5980 | 3.27E+06 |
| 0.64315 | 2.70E+06 | −7543611.5 | 2.3233 | −1.5472 | 3.05E+06 |
| 0.68043 | 2.42E+06 | −7825203.3 | 2.0965 | −1.4649 | 2.32E+06 |
| 0.72003 | 2.08E+06 | −8776458.4 | 1.8880 | −1.3861 | 1.01E+06 |
| 0.76035 | 1.69E+06 | −9542548.0 | 1.7046 | −1.3137 | −2.93E+05 |
| 0.80020 | 1.34E+06 | −8608247.4 | 1.5472 | −1.2490 | −2.40E+05 |
| 0.84037 | 1.02E+06 | −7878787.9 | 1.4085 | −1.1897 | −2.69E+05 |
| 0.88064 | 7.22E+05 | −7192192.2 | 1.2862 | −1.1355 | −2.59E+05 |
| 0.92030 | 4.55E+05 | −6601010.1 | 1.1797 | −1.0866 | −2.58E+05 |
| 0.96014 | 2.08E+05 | −6040201.0 | 1.0846 | −1.0415 | −2.24E+05 |
| 0.98005 | 9.74E+04 | −5570065.3 | 1.0411 | −1.0204 | 6.60E+03 |

Table 1. The relative radius, temperature, temperature gradient, gravitational acceleration, gravitational potential, and the gravitational temperature, of which all the source data are from the BS2005 standard solar model.

Eq.(5) introduces one new heat conduction mechanism into the astrophysical systems, which is associated with the whole nonequilibrium structure of the system and is mainly driven by the self-gravity to some extent. So its characteristic scale must be much more than the one of the conventional heat conduction taking place in the ground laboratory whose size is about tens of meters. It may be useful for the research on the thermal process and the thermodynamic evolution of the astrophysical systems.

Next, with the aid of Eq.(3), we will study the gravitational temperature distribution in the solar interior under the assumption of spherical symmetry. With the BS2005 standard solar model data, which are in excellent agreement with the measurement results of the helioseismology, we can calculate the gravitational temperature at different radius in the solar interior. The calculated results are listed in the Table 1 (The similar calculation has been done in Ref [15] for different purpose.). The value of the temperature constant $T_0$ in Eq.(4) is determined by the boundary condition at the top of the solar convection layer where the gravitational temperature should be equal to 6600K. The distribution curve along the solar radius of the gravitational temperature is drawn in Fig 2.

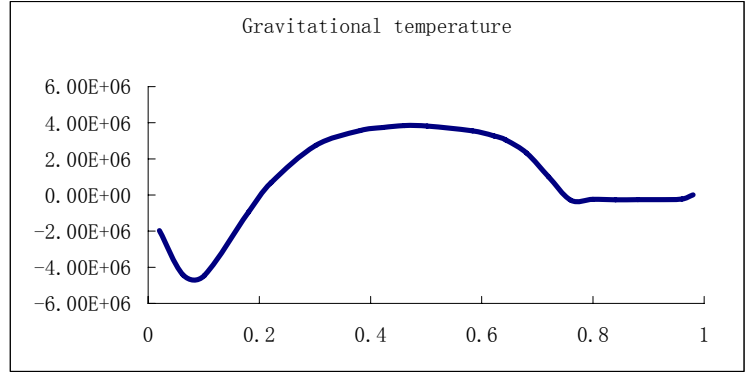

Fig.2. Distribution of the gravitational temperature with the relative solar radius $r/R_\odot$.

Firstly, from Table 1 and Fig.2 we can find that the gravitational temperature is negative in some regions such as the solar core and the convection layer. This is not strange, because the negative property of this generalized temperature is induced by the negative gravitational potential, now that we set the potential zero point being infinite far away. So the gravitational temperature reveals the kinetic energies and potential energies of the particles consisting of the self-gravitating system. That is why the gravitational thermal capacity defined dependent on the generalized temperature is possible to be positive [11].

Secondly, one can also see that the gravitational temperature in the solar interior is inhomogeneous along the radius. According to Eq.(5), this implies the gravitational heat conduction would appear in the solar interior. Obviously, the distribution curve in Fig 2 can be roughly divided into three parts. The first part is the region from the solar center to the position of $0.24R_\odot$, where $R_\odot$ is the solar maximum radius. The second part is from the $0.24R_\odot$ to the $0.74R_\odot$ and the third part is from the $0.74R_\odot$ to the solar surface. It is very apparent that these three parts are corresponding to the solar core region, radiation region and the convection region, respectively. Just in this sense we can say the hierarchical structure in solar interior is just the thermodynamic structure.

We don't have enough information to obtain the gravitational heat conduction coefficient in Eq.(5), so we can not calculate the exact value of the gravitational heat flux in different region of the sun. Even so, we can roughly analyze the directions and magnitudes of these gravitational heat fluxes. Here, we assume the whole sun is stable, so the total energy flux from the solar center to the surface should be a constant vector. Also we think that there are three mechanisms of energy transfer in the solar interior: radiation, gravitational heat conduction and convection.

In the solar core region, one notices that, in the region from about $0.08R_\odot$ to $0.24R_\odot$, the gravitational temperature gradient is positive. This indicates that the direction of the gravitational heat flux in this region points to the solar center. So there must be one huge radiation flux, or other energy flux, whose directions point to the solar surface, to reverse this. In the radiation region, the gradient of gravitational temperature trends to zero, so the main way of transferring energy in this region is the radiation, which is in agreement with the theory of standard solar model. In the convection region, the gradient of gravitational temperature is exactly zero; therefore there is not any gravitational heat conduction in this region. In this sense we insist such view point that the convection process is adiabatic [16]. Due to the zero heat flux, the main mechanism of energy transfer in the convective region is convection. There is no doubt that the radiation in this region is very weak, so the convection flux is very close to the value of the total energy flux in the solar interior.

In the view of thermodynamics, the main reason giving rise to the convection is the accumulation of large amount of heat at the bottom of the convection layer. Perhaps, this can be explained by the high gradient of gravitational temperature near the border of radiation region to the convection region. Due to the high gravitational heat flux in this juncture region, vast amount of heat accumulates here in each very short time. The convection layer has not enough time to take away the heat only by the conduction mechanism, so the convection motion takes place.

In summary, through the analysis to Tsallis equilibrium state of a self-gravitating system, we propose that there is not any heat conduction in such system. Therefore, at this state one state parameter called gravitational temperature can be defined. This definition is carried out according to Eq.(3) and the homogeneity of the nonextensive parameter $Q$. When the system is not at the Tsallis equilibrium state, the gravitational temperature is inhomogeneous and the gravitational heat conduction would appear.

According to the temperature duality [10], in Eq.(4) $T$ is the inverse of Lagrange multiplier, which is related to the collision effects of the molecules, and $T_g$ is the generalized temperature which is related to the nonextensivity of the system. The Lagrange temperature $T$ can be measured in physics; however the latter can not be measured in physics. In this sense, the Lagrange temperature is identical to the conventional thermodynamic temperature.

Generally speaking, Eq.(3) always holds no matter whether the system is at the Tsallis equilibrium or not, and no matter whether the parameter $Q$ is homogeneous or not. That is, the inhomogeneous gravitational temperature is characterized by the inhomogeneity of the parameter $Q$. This representation is convenient for the calculation of $Q$ by applying Eq.(3), and it is related to the generalized assumption of local equilibrium. We can use another representation: setting the parameter $Q$ always being homogenous as long as the system is at the hydrostatic equilibrium, when the system is not at the Tsallis equilibrium, Eq.(3) does not hold. Therefore, one has

$$\nabla T_g = \nabla T + \frac{Qm\nabla\varphi}{k}. \tag{6}$$

Within this representation Eq.(6), the physical meaning of the gravitational heat flux is very clear. If the value of Eq.(6) is positive, this means that the contribution of gravitational potential is more than the one of the (conventional) thermodynamic temperature. This leads to at least two results: the system absorbs some amount of heat from the environment and at the same time the system would expand a little (the virial theorem [17] holds).

On the contrary, if the value of Eq.(6) is negative, this implies the contribution of the thermodynamic temperature is beyond the one of the potential. There are also at least two results.

The system sends out some heat to the environment and meanwhile the system would shrink a little. Therefore, the gravitational heat flux is actually the residual effect of gradients. Ordinarily, we adopt the first representation, because it is convenient to calculate the parameter $Q$ with Eq.(3).

With the data of standard solar model BS2005 and with the assumption of spherical symmetry, we calculate the gravitational temperature in the solar interior. Then the distribution curve of the gravitational temperature along the radial direction is drawn. According to the gradient of gravitational temperature, this curve can be divided into three parts, which are corresponding to the solar core region, radiation region and the convection region, respectively. The conclusion can be drawn that, therefore, the solar hierarchical structure is the thermodynamic structure.

In the solar core, the gradient of gravitational temperature at the region from about $0.08R_\odot$ to $0.24R_\odot$ is positive with a large slope, showing the gravitational heat flux points to the solar center. Now that the whole sun is stable and therefore the energy flux from the center to the solar surface is a constant vector, it is doubtless that there exists a huge radiation flux or other energy flux there to balance the large heat flux.

The gradient of gravitational temperature in radiation region tends to zero, so the radiation plays an important role in this region. Due to the exact zero gradient of gravitational temperature, the main mechanism of energy transfer in the convection region is the convection motion. The thermodynamic motivation of the convection motion is derived from the juncture of the radiation and convection regions, where the gradient of the gravitational temperature is so large that there is a large amount of heat accumulation at the bottom of the convection layer. The heat accumulation amplifies the fluctuation of velocity, produces the macroscopic motion of fluid element and starts the convection motion.

In the whole paper, we always assume that $Q=1-q>0$, which is reasonable [9]. In view of this, Eq.(1) shows the heat cut and the non-ergodicity in the velocity space. Also according to this, the gravitational temperature in Eq.(4) can be negative, now that the potential is negative. According to the scalar form of Eq.(3),

$$Q = -k \frac{dT/dr}{mg}, \qquad (7)$$

$Q>0$ reveals the character of temperature distribution of the self-gravitational system.

In a word, the gravitational heat conduction, whose characteristic scale is the same as the one of the self-gravitating system, is firstly introduced into the astrophysical systems. This heat conduction mechanism is different from the one taking place ordinarily in the ground laboratory with a size of dozens of meters. The latter is driven by the conventional thermodynamic temperature gradient; therefore it only appears in the laboratory system. That is to say, the thermodynamic temperature gradient can not drive the large-scale heat conduction, because the self-gravity forbids this. Only the gradient of gravitational temperature, which contains the action of potential gradient, can take this task.

***

This work is supported by the National Natural Science Foundation of China under Grant No.11175128 and by the Higher School Specialized Research Fund for Doctoral Program under grant No 20110032110058.